\documentclass[a4paper,fleqn,11pt]{cas-sc}

\usepackage{setspace}
\usepackage{lineno}
\usepackage{multirow}
\usepackage{amsmath}
\usepackage{amssymb}
\usepackage{bm}
\usepackage{graphicx}
\usepackage{enumerate}
\usepackage{xcolor}
\usepackage[numbers,sort&compress]{natbib}

\newcommand{\blue}[1]{{\color{blue}{#1}}}

\newcommand{\sref}[1]{\blue{Section}~\ref{#1}}

\newcommand{\fref}[1]{\blue{Figure}~\ref{#1}}

\newcommand{\tref}[1]{\blue{Table}~\ref{#1}}

\begin{document}
\let\WriteBookmarks\relax
\def\floatpagepagefraction{1}
\def\textpagefraction{.001}

\shorttitle{Early Identification of Information Propagation Sources}

\shortauthors{Wen T. et~al.}

\title{\Large{EAGLE: Early Identification of Information Propagation Sources through Generalized Bayesian Inference in Large-scale Social Networks}}

\author[1,2]{Tao Wen}
\author[1,2]{Yu-wang Chen}
\cormark[1]
\author[3]{Jie Zhao}
\author[4]{Changchun Pan}
\author[1]{Tahir Abbas Syed}
\author[5]{Yong-Yeol Ahn}
\author[3,6]{Kang Hao Cheong}

\affiliation[1]{organization={Centre for AI and Decision Sciences, Alliance Manchester Business School, The University of Manchester}, city={Manchester}, postcode={M15 6PB}, country={UK}}
\affiliation[2]{organization={The Alan Turing Institute}, city={London}, postcode={NW1 2DB}, country={UK}}
\affiliation[3]{organization={Division of Mathematical Sciences, School of Physical and Mathematical Sciences, Nanyang Technological University}, city={Singapore}, postcode={S637371}, country={Singapore}}
\affiliation[4]{organization={School of Electronic Information and Electrical Engineering, Shanghai Jiao Tong University}, city={Shanghai}, postcode={200240}, country={China}}
\affiliation[5]{organization={School of Data Science, University of Virginia}, city={Charlottesville}, postcode={VA 22902}, country={USA}}
\affiliation[6]{organization={College of Computing and Data Science, Nanyang Technological University}, city={Singapore}, postcode={S639798}, country={Singapore}}

\cortext[label1]{Corresponding author: Yu-wang Chen (Yu-wang.Chen@manchester.ac.uk), Centre for AI and Decision Sciences, Alliance Manchester Business School, The University of Manchester}

\begin{abstract}
    In the era of infodemics, rapidly pinpointing information propagation sources is essential for mitigating negative impacts and enhancing societal resilience. This study presents EAGLE (\underline{E}arly identific\underline{A}tion of propagation sources through \underline{G}enera\underline{L}ized Bayesian inf\underline{E}rence), a novel and effective framework for early source identification in large-scale social networks, which integrates evolutionary computation to strategically deploy sparse observers with generalized Bayesian inference to aggregate observational data of varying credibility. Unlike existing approaches, our method relies on limited input from human observers and explicitly incorporates the uncertainty and noise of the data. Comprehensive evaluations across diverse networks and propagation scenarios demonstrate that our approach consistently surpasses state-of-the-art models, without relying on assumptions about propagation characteristics or user attributes. Utilizing real-world datasets with empirical propagation paths from Twitter, our method is among the first capable of identifying potential sources in large-scale networks using limited observers, demonstrating its broad applicability. These findings highlight the effectiveness of generalized Bayesian inference in uncovering propagation dynamics and equip organizations and policymakers with critical tools to proactively address threats to digital trust and security.
\end{abstract}


\maketitle

\section{Introduction}\label{Sec_introduction}

As our society becomes more connected, both physically and electronically, the increased connectedness brings both immense benefits and tough challenges to humanity~\cite{jusup2022social,pedreschi2025human}. Online social platforms, for instance, eliminate spatial-temporal constraints of traditional communication, fostering instantaneous global interactions~\cite{avalle2024persistent,eiben2023parameterized}. However, the same characteristics that allow for rapid information exchange also facilitate the uncontrolled spread of misinformation and disinformation, posing a major challenge for human societies~\cite{loomba2021measuring,KNOP2026104563,CALDARELLI20261}. Similarly, the interconnectedness of modern transportation networks, while pulling the whole world closer, has escalated the risk of disease outbreaks and pandemics, posing severe threats to both health and economic stability~\cite{douven2021mis,cheong2022alternating}.

The rapid and uncontrolled spread of misinformation highlights the critical need for effective monitoring and early intervention in complex networks, as effective management of these processes is vital for maintaining societal stability and trust~\cite{ji2023signal}. Researchers have approached this challenge from multiple perspectives, including identifying influential spreaders~\cite{wen2024eriue,liu2024game}, analyzing network characteristics~\cite{meena2023emergent,zhao2023obfuscating}, and facilitating group decision-making under uncertainty~\cite{wen2024formulating,fang2022integrating,berenbrink2026opinion}. Among these efforts, localizing the propagation source is particularly crucial, as it forms the foundation for timely and targeted interventions~\cite{wang2022rapid,jiang2016identifying}. 

A wide range of methods have been proposed for source identification, including approaches based on maximum likelihood estimation, machine learning, and nonlinear dynamics. These approaches can broadly be grouped by their observation paradigm, including complete observation, snapshot observation, and sensor-based observation~\cite{cheng2025efficient,zhao2024mase,liu2023diffusion,cheng2024heuristic}. Complete observation methods assume full access to the exact states of all nodes at a given time, allowing direct inference of the source. For example, rumor centrality~\cite{shah2011rumors} identifies sources in tree-like networks by considering the number of distinct propagation paths, while the local rumor center method~\cite{dong2013rooting} narrows the search space by limiting the set of potential sources. Under the Susceptible-Infective (SI) model, methods such as minimum description length~\cite{prakash2012spotting} and dynamic age~\cite{fioriti2014predicting} leverage spectral properties of the Laplacian matrix. Despite their effectiveness, such methods are rarely feasible in practice, given the substantial data, computational, and knowledge requirements, particularly in large-scale and dynamic systems. 

Snapshot-based approaches relax these requirements by relying on partial observations of network states at a specific time. These methods typically require detailed knowledge of node attributes, such as influence intensity and susceptibility, and often assume diffusion models like SI or Susceptible–Infective–Recovered (SIR)~\cite{wang2022invertible,hou2023sequential}. The Jordan center method~\cite{zhu2014information}, for instance, identifies the source by considering the optimal sample path that most likely results in the observed snapshot, while message passing~\cite{lokhov2014inferring} infers the most likely source via iterative probability updates. More recently, temporal-sequence based graph attention models~\cite{hou2023sequential} apply inductive learning to estimate the influence probability between users and distinguish the importance of sources without prior assumptions. Despite their sophistication, these methods often depend heavily on assumed diffusion models and tree-like structures, limiting their applicability to real social networks.

To overcome the limitations of complete and snapshot observations, observer-based (or sensor-based) approaches provide a more feasible solution by collecting data from strategically placed or randomly deployed observers, which reduces the amount of required information~\cite{zhao2025enhanced,wang2022rapid,zhao2024mase}. Observers typically record the arrival time and direction of information from their neighbors. A seminal example is the Gaussian Estimator (GE)~\cite{pinto2012locating}, which assumes Gaussian-distributed propagation delays in tree-like networks. To improve computational efficiency, the gradient maximum likelihood algorithm~\cite{paluch2020enhancing} limits the search to high-probability candidates. The Direction-Induced Search based Gaussian Estimator (DISGE)~\cite{yang2020locating} further integrates directional information to refine source inference. In addition, Bayesian source estimators~\cite{altarelli2014bayesian} employ iterative updates of belief propagation equations to infer the most probable source, though these too are primarily applicable to tree-like topologies. Beyond these, several heuristic or simulation-based techniques have been proposed~\cite{wen2021fractal,wen2024parrondo}, such as label propagation~\cite{cheng2022path}, effective distance~\cite{lalou2016least,zhao2023early}, and Monte Carlo~\cite{agaskar2013fast} methods, each with specific strengths and weaknesses.

Despite substantial progress, existing methods face notable challenges when applied to real social media environments. First, many rely on strong assumptions about network topology (tree-like structures), propagation models (SI/SIR), or statistical properties such as propagation delays, which are rarely available in real-world social media~\cite{prakash2012spotting,fioriti2014predicting,wang2022rapid,zhu2014information}. Second, most observer-based methods neglect optimal observer deployment, often placing sensors randomly or by degree centrality, which reduces localization accuracy despite similar resource costs~\cite{paluch2020enhancing,cheng2022path,pinto2012locating}. Third, these methods generally assume post-propagation data collection, requiring all observers to be exposed before inference, which delays intervention and allows further spread~\cite{wang2020locating,lalou2016least}. Fourth, they typically treat all observers equally, failing to account for heterogeneous reliability, noise, or structural positioning~\cite{pinto2012locating,yang2020locating}. Finally, traditional probability-based frameworks struggle to capture epistemic uncertainty: when multiple candidates are indistinguishable, they collapse belief into equal probabilities rather than explicitly representing uncertainty. This fundamentally limits inference accuracy under limited knowledge. Addressing this challenge is essential for developing robust and generalizable source localization frameworks that function reliably under the incomplete and uncertain conditions characteristic of real-world information ecosystems.

Therefore, accurately inferring the propagation source in large-scale social networks remains challenging when the spread rate is low, the available observations are sparse, and no prior knowledge of the propagation dynamics and user attributes is available. To address these concerns, we propose EAGLE (\underline{E}arly identific\underline{A}tion of propagation sources through \underline{G}enera\underline{L}ized Bayesian inf\underline{E}rence), a novel and generalizable framework for early and robust source localization. EAGLE first minimizes observational blind spots through the strategic deployment of observers, identified via a discrete moth-flame optimization algorithm with a dedicated fitness function~\cite{mirjalili2015moth}. This optimization ensures a balanced trade-off between coverage and redundancy, enabling efficient information capture across large-scale networks. The framework requires only essential propagation data, including the timing and direction of observed propagation events, from a limited number of exposed observers. This design allows EAGLE to operate effectively even in early-stage propagation scenarios and when detailed knowledge of user attributes and diffusion parameters is unavailable. 

Crucially, EAGLE represents the likelihood of a candidate being the source using a belief function~\cite{shafer1976mathematical} rather than a conventional probability function. This enables the explicit representation of uncertainty and indistinguishability among candidate sources. The generalized Bayesian inference framework~\cite{yang2013evidential,yang2025maximum} then integrates the likelihoods derived from various observer combinations in an explainable and theoretically grounded manner, offering a flexible, interpretable, and generalizable foundation for source localization. To further enhance reliability, EAGLE assigns distinct credibility levels to each combination of exposed observers, determined by their structural advantages within the network topology. This weighting mechanism mitigates the impact of noise and incomplete data, thereby improving inference robustness and accuracy. EAGLE's performance has been validated across diverse real-world networks and varied propagation patterns, consistently outperforming state-of-the-art (SOTA) methods in both accuracy and efficiency. Importantly, EAGLE demonstrates robust performance in real Twitter networks with about 100,000 users, accurately identifying sources even when only 20 to 120 observers provide valid information. To the best of our knowledge, this represents the first validation of source localization on the largest real social network with reconstructed propagation paths, where EAGLE achieves both a significantly reduced inferred source set and a marked improvement in F-score, outperforming all comparative SOTA models under the same conditions.

The main contributions of this study are summarized as follows.
\begin{enumerate}[1)]
    \item A discrete moth–flame optimization algorithm is employed to strategically deploy a sparse set of observers via a dedicated fitness function, achieving a balanced trade-off between coverage and redundancy.
    \item The belief function is used instead of traditional probability measures to quantify the likelihood of each candidate being the source, enabling explicit modeling of uncertainty and indistinguishability among candidates. The method requires only the timing and direction of propagation arrivals at a limited number of exposed observers, allowing for early-stage identification with minimal human input.
    \item Sparse and noisy observational data are integrated through a generalized Bayesian inference model, where each observer combination is weighted according to its structural credibility. This mechanism enhances localization accuracy by systematically accounting for noise, uncertainty, and ignorance in the inference process.
    \item EAGLE reformulates source localization as a multi-criteria decision-making problem, providing a transparent, interpretable, and extensible framework that can incorporate additional information sources and reasoning mechanisms in future developments.
\end{enumerate}

The remainder of this paper is structured as follows. \sref{sec.proform} defines the problem. \sref{sec.method} presents the EAGLE approach, detailing its integration of evolutionary computation and generalized Bayesian inference. The experimental setup, comparative analysis with SOTA methods, and results are described in \sref{sec.experiment}, highlighting the effectiveness and applicability of EAGLE. Finally, \sref{sec.discussion} discusses key findings and contributions.


\section{Problem Formalization} \label{sec.proform}

Consider a general undirected and unweighted social network $\mathcal{G}\left( \mathcal{N}, \mathcal{E} \right)$, where $\mathcal{N}$ and $\mathcal{E}$ denote the set of nodes and pairwise edges, respectively. An information propagation (or social contagion) process originates from an unknown source node $s^* \in \mathcal{N}$. The objective is to infer this source based on limited and sparse observations.

\begin{figure*}[!tb]
\begin{center}
\includegraphics[width=\textwidth]{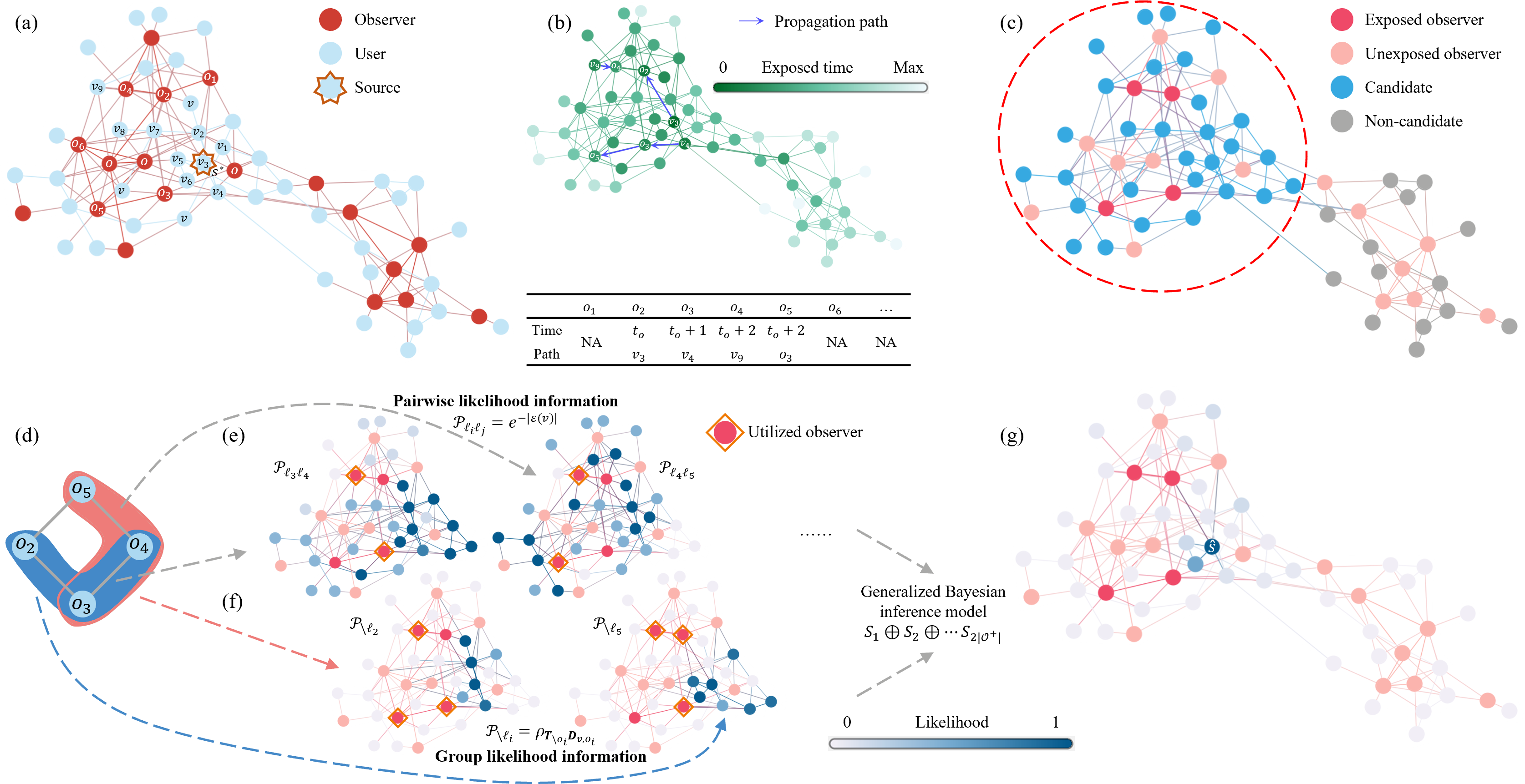}
\end{center}
\caption{\textbf{Overview of the EAGLE framework.} \textbf{(a)} Example of a dolphin social network, where node $v_{3}$ represents the true source of propagation, $s^*$. Observers are strategically deployed using an optimization approach. \textbf{(b)} Visualization of the propagation process under a heterogeneous SI epidemic model, with node color denoting exposure time. Observations from four observers, including both timing and path of the propagation, are used to infer the source at an early stage, as shown in the accompanying table. Paths from these observers are highlighted in the network. \textbf{(c)} Classification of exposed and unexposed observers, candidates, and non-candidates. \textbf{(d)} Illustration of four exposed observers used for comparative analysis of data, including \textbf{(e)} pairwise likelihood evaluations between two observers and \textbf{(f)} group likelihood assessments among a set of observers, arranged in a ring to represent interconnections. The color gradients denote the inferred likelihood of each candidate being the true source. \textbf{(g)} Final likelihood estimation for each candidate. The source $\hat{s}$ identified by the EAGLE approach correctly matches the true source $s^*$.}
\label{fig:Framework}
\end{figure*}

To this end, we deploy a small set of observers $\mathcal{O} \triangleq \{o_1, \ldots, o_i, \ldots, o_{|\mathcal{O}|}\} \subset \mathcal{N}$ strategically within the network, as illustrated in \fref{fig:Framework}(a). Since deploying and maintaining observers incurs cost, their number is constrained to be a small fraction of the network size, $\left | \mathcal{O} \right | = \xi \left | \mathcal{N} \right |$ with $\xi \ll 1$. Importantly, observer deployment is conducted prior to any propagation event, meaning that the configuration is independent of the specific content or initiating user. We also exclude trivial cases where the true source belongs to the observer set, $s^* \notin \mathcal{O}$. As information spreads along network edges, each observer $o_i$ records two essential features of the first received signal, including the time of arrival $t_i \in \mathbb{N}^0$ and the neighbor $s_i \in \mathcal{N}$ from whom the information was received. Each observation is thus represented as $\ell_i = \{o_i, t_i, s_i\}$, as shown in \fref{fig:Framework}(b). The actual start time of propagation $t_s$ is unknown and independent of any recorded $t_i$. We distinguish between exposed and unexposed observers, denoted respectively as $\mathcal{O}^+$ and $\mathcal{O}^-$ with $\mathcal{O} = \left \{ \mathcal{O}^+, \mathcal{O}^- \right \}$ (\fref{fig:Framework}(c)). Because the source needs to be identified at an early stage, the number of exposed observers is typically much smaller than the total number deployed, $\left | \mathcal{O}^+ \right | = \varphi \left | \mathcal{O} \right |$ with $\varphi \ll 1$. Only data from exposed observers $o_i \in \mathcal{O}^+$ are used for inference. To further reduce computational complexity, the set of candidate sources $\mathcal{C}$ is restricted to nodes proximal to the exposed observers, highlighted by the red box in \fref{fig:Framework}(c). 

The problem of source localization can therefore be formalized as estimating the most probable source $\hat{s}$ given the observed data $\{\mathcal{G}, \mathcal{O}^+\}$:
\begin{equation}
    \hat{s} = \arg \max_{v \in \mathcal{C}} \mathbf{P} \left( s^* = v \mid \mathcal{O} \right),
\end{equation}
where $\mathbf{P} \left( s^* = v \mid \mathcal{O} \right)$ denotes the conditional probability that user $v$ is the true source given all available observer reports.

\section{Methodology} \label{sec.method}

Our EAGLE approach encompasses two main processes: strategic deployment of observers using a discrete moth-flame optimization algorithm in \fref{fig:Framework}(a), and localization of the propagation source through a generalized Bayesian inference model in \fref{fig:Framework}(d) -- (g). 

\subsection{Observers Deployment}

\textit{Fitness Function.} The primary objective of deploying observers in a network is to monitor as many candidate users as possible and to cross-observe each candidate's propagation behavior through multiple observers. In this work, observers are assumed to be able to only observe the behavior of their one-hop neighbors directly. Hence, for each candidate user $c_j$, it is critical that their one-hop neighbor set includes a sufficient number of observers. The number of observers who can observe a given candidate user $c_j$, is represented by $\left | \Gamma_{c_j}^\mathcal{O} \right | = \sum_{o_i \in \mathcal{O}} \mathbf{1}_{o_i \in \Gamma_{c_j}}$, where the indicator function $\mathbf{1}_{o_i \in \Gamma_{c_j}}$ is defined as,
\begin{equation}
    \mathbf{1}_{o_i \in \Gamma_{c_j}} := \left\{\begin{matrix}
    1, & \text{if } o_i \in \Gamma_{c_j} \\ 
    0, & \text{if } o_i \notin \Gamma_{c_j}
    \end{matrix}\right.,
\end{equation}
and $\Gamma_{c_j}$ denotes the neighborhood set of user $c_j$.

If observers are clustered around a limited number of candidates, the propagation behaviors of others may not be adequately monitored, resulting in significant information loss within the network. To avoid this, the distribution of observers around each candidate is regulated using weight coefficients $\omega_l$, where $\omega_l$ represents the weighting factor for the scenario in which there are $l$ observers in the one-hop neighborhood of a specific user, defined as
\begin{equation}
    \omega_l = \left\{\begin{matrix}
    -\lambda, & \text{if } l = 0  \\ 
    \exp\left( -\frac{\lfloor \bar{k} \rceil-l}{2} \right), & \text{if } 0 < l < \lfloor \bar{k} \rceil \\
    1, & \text{if } l \geq \lfloor \bar{k} \rceil
    \end{matrix}\right.,
\end{equation}
where $\lambda \in \mathbb{N}^+$ is a penalty applied when there are no observers ($\left | \Gamma_{c_j}^\mathcal{O} \right | = 0$), $\bar{k}$ is the average degree of the network, and $\lfloor \cdot \rceil$ indicates the nearest integer rounding. This approach ensures that having observers exceeding $\lfloor \bar{k} \rceil$ does not receive additional weight, promoting a balanced distribution. 


The effectiveness of an observer set $\mathcal{O}$ is quantified by the fitness function $F(\mathcal{O})$, evaluated by
\begin{equation} \label{eq.fitness}
    F \left( \mathcal{O} \right) = \sum_{c_j \in \mathcal{C}}  \omega_l \times \mathbf{1}_{\left | \Gamma_{c_j}^\mathcal{O} \right | = l},
\end{equation}
where the goal is to maximize $F(\mathcal{O})$ to achieve the optimal observer deployment. Given the NP-hard nature of this optimization problem, especially in large-scale networks, we employ the discrete moth-flame optimization algorithm~\cite{mirjalili2015moth,wang2021identifying}, a swarm intelligence-based method, to determine the optimal set of observers $\mathcal{O}$. 
Specifically, the Moth-Flame Optimization (MFO) algorithm is inspired by the logarithmic spiral-like flight pattern of moths around a flame~\cite{mirjalili2015moth}. Originally designed for continuous search spaces, it updates the moth population using
\begin{equation}
    M_i^{T+1} = \left| F_j^T - M_i^T \right| \exp{\left( \lambda \right)} \cos{\left( 2 \pi \lambda \right)} + F_j^T,
\end{equation}
where $M_i$ and $F_j$ denote the positions of the $i$-th moth and $j$-th flame, respectively, $T$ denotes the iteration number, and $\lambda \in \left[ -1, 1 \right]$ is a random coefficient adjusting the spiral flight path. However, for applications like observer set deployment in networks -- a discrete space composed of user IDs -- a discrete version of MFO is required.

\textit{Population Initialization and Search Area Determination.} For the moth population at the $T$-th iteration, there are $\left| M \right |$ moths with seed set size $\left| \mathcal{O} \right |$, which is expressed as
\begin{equation}
    M^T = \begin{bmatrix}
    M_1^T \\ 
    \vdots \\ 
    M_{\left| M \right |}^T
    \end{bmatrix} = 
    \begin{bmatrix}
    m_{1, 1} & \cdots & m_{1, \left| \mathcal{O} \right |} \\ 
    \vdots & \ddots & \vdots \\ 
    m_{\left| M \right |, 1} & \cdots & m_{\left| M \right |, \left| \mathcal{O} \right |}
    \end{bmatrix}.
\end{equation}
Each moth's seed set, $M_i^0$, is initialized using a heuristic strategy based on node degree, selecting members according to $\vartheta \times k_i$, where $\vartheta \in \left[ 0, 1 \right]$ is a uniformly distributed random number. This approach ensures a higher probability of selecting highly connected nodes while still allowing for random lower-degree node selection. To ensure the involvement of the near-optimal solution in the initial population, one of the moths is specifically composed of the top \(\left| \mathcal{O} \right|\) highest-degree individuals.

For swarm intelligence algorithms, establishing a reasonable and effective search area is crucial for finding the optimal solution. A small search area may lead to falling into local optima, while a too large search area could be inefficient in finding the optimal solution. In social networks, highly connected users are more likely to effectively observe the propagation from a larger number of users. Hence, the top $\lfloor \kappa \times \left| \mathcal{O} \right | \rceil$ users with the highest degree are selected as the search area. 
Here, $\kappa = 4$ is chosen based on previous studies~\cite{wang2021identifying}, which have identified it as the appropriate range using the traditional greedy strategy.

\textit{Population Updating and Mutation.} The discrete nature of the search space requires adaptations for moth population updating. In this work, the local crossover process involves sorting the moth population by fitness and designating $M_1^T$ as the temporary optimal solution. Each moth's position, $M_i^T$, updates according to $x_{i, j}$ in $X_i^T = \left[ x_{i, 1}, x_{i, 2}, \cdots, x_{i, j}, \cdots, x_{i, \left| \mathcal{O} \right |} \right]$,
\begin{equation}
    x_{i, j} = \text{sigmoid} \left[ b_l + \left( b_u - b_l \right) \times \vartheta \right],
\end{equation}
where $x_{i, j}$ is generated by the sigmoid function with a random number $\vartheta \in [0,1]$. Hence, the independent variable is in the range of $\left[ b_l, b_u \right]$. In this work, $b_l = -5$ and $b_u = 5$ because saturation occurs when the independent variable is out of $[-5, 5]$. Its crossover location with the temporary optimal moth is determined by $P_i^T = \left[ p_{i, 1}, p_{i, 2}, \cdots, p_{i, j}, \cdots, p_{i, \left| \mathcal{O} \right |} \right]$, and
\begin{equation}
    p_{i, j} = \mathbf{1}_{x_{i, 1} > \gamma} \otimes \mathbf{1}_{x_{i, j} > \gamma},
\end{equation}
where $\gamma \in \left[ 0, 1 \right]$ is a random threshold and $\mathbf{1}$ is the indicator function that equals 1 only if the condition is true. As for the arithmetic symbol $\otimes$, it equals $1$ only if both operands are $1$, otherwise the result is $0$. Elements from $M_i^T$ undergo exchanges based on these crossover points, enhancing the solution diversity. For better-performing moths $M_i^T, i \leq \tau$, its $j$-th element undergoes a two-way exchange with the element in the temporary optimal moth $M_1^T$ when $p_{i, j} = 1$, and the $j$-th element of other moths $M_i^T, i > \tau$ only takes a one-way replacement from $M_1^T$ when $p_{i, j} = 1$, thereby retaining the performance of the best-known optimal moth for the next iteration $M_i^{T+1}$.

To prevent premature convergence, mutation occurs with probability $\varrho$ for high-performing moths $M_i^T, i \leq \tau$, replacing parts of their configuration with randomly selected individuals from the search area, where $\tau = 5$. Duplicate elements within a moth are replaced to maintain unique user representation in each solution set. 

\subsection{Information Collection}

In the deployed social network, only a limited subset of observers $\mathcal{O}^+$ 
are responsible for capturing and reporting propagation data. These exposed observers provide critical information for localizing the propagation source, including the timing and direction of the information received. Specifically, an observer $o_i \in \mathcal{O}^+$ contributes the tuple $\ell_i = \{o_i, t_i, s_i\}$, indicating that they received information from user $s_i$ at time $t_i$. Unexposed observers and other candidates, who either have not observed or are unable to observe the propagation behavior, do not contribute to the data collection. 

To reduce computational complexity, candidates for the source are first narrowed down to users closer to the exposed observers compared to those unexposed based on the shortest distance in the network (\fref{fig:Framework}(c)),
\begin{equation}
    \mathcal{C} \triangleq\left\{v \in \mathcal{N} \setminus \mathcal{O}: \bar{d}(v,\mathcal{O}^{+}) < \bar{d}(v,\mathcal{O}^{-})\right\},
\end{equation}
where $\bar{d}$ indicates the average distance of the shortest paths. The information from exposed observers is then combined in various manners to infer the potential propagation source, utilizing both pairwise and group arrangements of observers.

\begin{figure}[!tb]
\begin{center}
\includegraphics[width=0.4\textwidth]{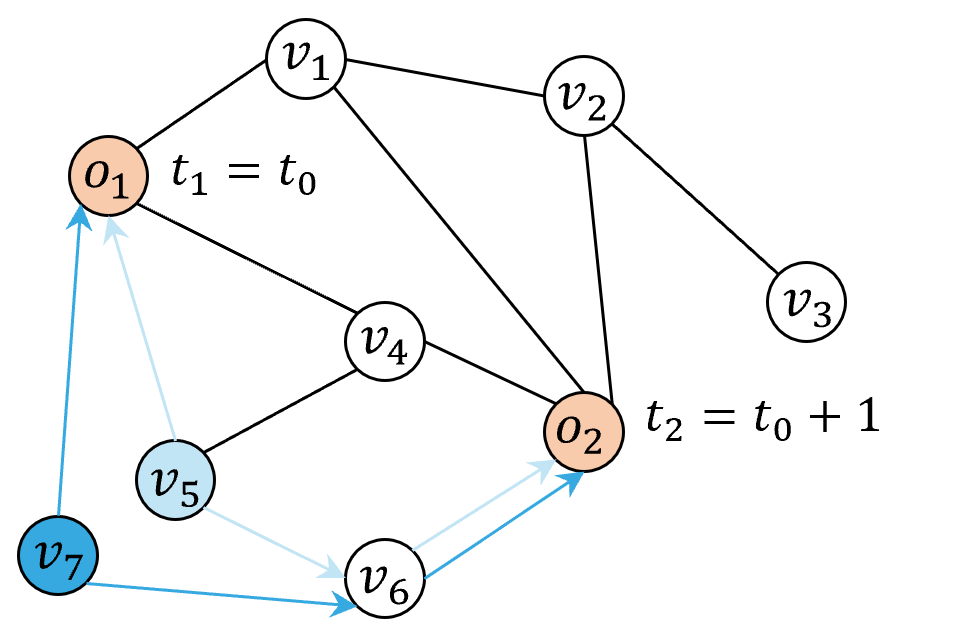}
\end{center}
\caption{Illustrative example of the observation process.}
\label{fig:ExampleFlow}
\end{figure}

\textit{Pairwise Likelihood Information.} 
The likelihood that a particular candidate is the source is determined based on the data from any two exposed observers. Under the extreme condition, where the probability of information propagating along each link is 1, the time at which an observer is informed should correspond directly to their shortest network distance from the source $s^*$, such that $d(s^*, o_i) / t_i = 1$. Hence, for two observers, the difference in the times they are informed should match the difference in the distance of their shortest paths from the source, $\left({d(s^*, o_i) - d(s^*, o_j)}\right)/\left({t_i - t_j}\right) = 1$. For example, in the network shown in \fref{fig:ExampleFlow} when the propagation probability is 1, the time difference between the two observers being informed is 1, thus, the possible source can only be $v_5$ or $v_7$, because the distance difference of the shortest paths from these sources to the two observers is 1, where the shortest propagation paths from $v_5$ and $v_7$ to the two observers are marked in different colors, respectively. 
In reality, although propagation probability along each edge would not be 1, we still think that the ratio is approximately 1. Therefore, the likelihood that a user $v$ is the source, as inferred by observers $o_i$ and $o_j$ and their observations $\ell_i$ and $\ell_j$, can be expressed by,
\begin{equation}
    \begin{gathered}
        \mathcal{P}_{\ell_i \ell_j} \left( v \right) = e^{-\left | \varepsilon\left( v \right) \right |}, \\
        \varepsilon\left( v \right) = 
            \left\{\begin{matrix}
                1 - \frac{d' \left( v, s_i \right) - d' \left( v, s_j \right)}{t'_i - t'_j}, & \text{if } t'_i \neq t'_j \\ 
                d' \left( v, s_i \right) - d' \left( v, s_j \right), & \text{if } t'_i = t'_j
            \end{matrix}\right.,
    \end{gathered}
\end{equation}
where $d'$ and $t'$ are the normalized values of distances and times. The pairwise likelihood of each candidate user provided by $\left\{ o_i, o_j \right\}$ is thus expressed as $\mathbf{G}_{\ell_i \ell_j} = \left\{ \mathcal{P}_{\ell_i \ell_j} \left( v_1 \right), \mathcal{P}_{\ell_i \ell_j} \left( v_2 \right), \cdots, \mathcal{P}_{\ell_i \ell_j} \left( v_{\left | \mathcal{C} \right |} \right) \right\}$. Pairs of observers form a ring structure, as shown in \fref{fig:Framework}(d). 
This complete comparative information chain allows each observer to cross-validate information and analyze the potential source comparatively with any other observer, 
inferring the propagation source more reasonably. In addition, there are $|\mathcal{O}^+|$ pieces of pairwise information from the $|\mathcal{O}^+|$ exposed observers.

\textit{Group Likelihood Information.} 
A group of exposed observers can collectively evaluate the likelihood of candidate sources based on a consensus of their observations. The assumption is that observers closer to the source will likely report earlier, and thus observation times and minimum hop distances to the source would be correlated. 
For a group of observers, $\mathcal{O}^+ \setminus o_i = \left\{ o_1, \cdots, o_{i-1}, o_{i+1}, \cdots, o_{\left | \mathcal{O}^+ \right |} \right\}$, they can provide the sets of times and distances for inference, 
\begin{equation}
    \begin{split}
        \mathbf{T}_{\setminus o_i} &= \left\{ t_1, \cdots, t_{i-1}, t_{i+1}, \cdots \right\}, \\ 
        \mathbf{D}_{v, \setminus o_i} &= \left\{ d \left( v, s_1 \right), \cdots, d \left( v, s_{i-1} \right), d \left( v, s_{i+1} \right), \cdots \right\}, 
    \end{split}
\end{equation}
where observer $o_i$'s data is excluded. The likelihood of a candidate $v$ being the source estimated by this group of observers can be determined by 
\begin{equation}
    \begin{split}
        \mathcal{P}_{\setminus \ell_i} \left( v \right) &= \rho_{\mathbf{T}_{\setminus o_i} \mathbf{D}_{v, \setminus o_i}} = \frac{cov\left(\mathbf{T}_{\setminus o_i}, \mathbf{D}_{v, \setminus o_i}\right)}{\sigma_{\mathbf{T}_{\setminus o_i}} \sigma_{\mathbf{D}_{v, \setminus o_i}}} \\ 
        & = \frac{\mathbb{E}\left [ \left ( \mathbf{T}_{\setminus o_i} - \mu_{\mathbf{T}_{\setminus o_i}} \right ) \left ( \mathbf{D}_{v, \setminus o_i} - \mu_{\mathbf{D}_{v, \setminus o_i}} \right ) \right ]}{\sigma_{\mathbf{T}_{\setminus o_i}} \sigma_{\mathbf{D}_{v, \setminus o_i}}},
    \end{split}
\end{equation}
where $\rho$ denotes the Pearson correlation coefficient, with $\mu$ and $\sigma$ representing the mean and standard deviation, respectively. Therefore, the group likelihood information of candidate users provided by $\mathcal{O}^+ \setminus o_i$ can be expressed as $\mathbf{G}_{\setminus \ell_i} = \left\{ \mathcal{P}_{\setminus \ell_i} \left( v_1 \right), \mathcal{P}_{\setminus \ell_i} \left( v_2 \right), \cdots, \mathcal{P}_{\setminus \ell_i} \left( v_{\left | \mathcal{C} \right |} \right) \right\}$. Due to the stochastic nature of the propagation process, information can spread across regions via long-range ties~\cite{park2018strength} with extremely low probability, potentially misleading the observer $o_i$. In addition, observers themselves may experience faults~\cite{liu2023evidential,liu2024online}, leading to unreliable observational data, including inaccuracies in time $t_i$ and direction $s_i$. To mitigate uncertainty and unknowns in this process, each individual observer $o_i$ is removed once in the group to account for potential unreliability, and the remaining observers are then used to infer the propagation source, thereby enhancing the accuracy of the inference. This results in $|\mathcal{O}^+|$ pieces of group information involved in the inference process. Moreover, at least four exposed observers are necessary to consider group likelihood information effectively; otherwise, it would degenerate into pairwise likelihood information.

Collectively, both pairwise and group likelihood information contribute to a comprehensive set of likelihood scores for each candidate, denoted as $\mathbf{G}_l$ to simplify the expression for clarity,
\begin{equation}
    \mathbf{G}_{l} = \left\{ \mathcal{P}_l \left( v_1 \right), \mathcal{P}_l \left( v_2 \right), \cdots, \mathcal{P}_l \left( v_{\left | \mathcal{C} \right |} \right) \right\},
\end{equation}
where $l \in \left\{ \cdots, \ell_i \ell_j, \ell_j \ell_k, \cdots, \setminus o_1, \cdots, \setminus o_i, \cdots, \setminus o_{ \left | \mathcal{O}^+ \right | } \right\}$. 
In the subsequent process, all pairwise and group likelihood information will be combined with varying credibility based on the structural characteristics of the observers, thereby locating the propagation source reasonably.

\subsection{Source Inference and Localization}

In the example shown in \fref{fig:ExampleFlow}, the information collected by observers cannot distinguish whether the true source is $v_5$ or $v_7$, but it can conclusively eliminate all other candidates. A naive probabilistic approach would assign equal probabilities, $p(v_5) = p(v_7) = 0.5$ and $p(v_1) = p(v_2) = \cdots = 0$, yet this fails to capture the uncertainty between $v_5$ and $v_7$, that is, we cannot ensure they have the same likelihood $p(v_5) = p(v_7)$. To represent such uncertainty, we employ belief functions from Dempster–Shafer evidence theory~\cite{shafer1976mathematical}, setting $m(\{v_5, v_7\}) = 1$ to denote that the source certainly lies within $\{v_5, v_7\}$ but remains unspecified between them, where $m(\cdot)$ indicates the degree of support assigned to a proposition.

Specifically, the candidate set $\mathcal{C} = \{C_1, C_2, \cdots, C_{\left| \mathcal{C} \right|}\}$ is a collection of mutually exclusive and collectively exhaustive propositions about potential sources, such that $C_i  \cap C_j = \emptyset$ for $\forall i \neq j \in \{ 1, 2, \cdots, \left| \mathcal{C} \right|\}$ and $\emptyset$ denotes an empty set. The candidate set $\mathcal{C}$ is also referred to as the frame of discernment. The belief function (or basic probability assignment) is a function $m: 2^\mathcal{C} \rightarrow [0, 1]$, satisfying,
\begin{equation}
    m_\emptyset = 0, \quad \sum_{\Omega \subseteq \mathcal{C}} m_\Omega = 1,
\end{equation}
where $m_\Omega$ quantifies the degree of belief that the true source lies within subset $\Omega \subseteq \mathcal{C}$. Here, $2^\mathcal{C}$ or $P \left( \mathcal{C} \right)$ denotes the power set of $\mathcal{C}$, containing all $2^{\left| \mathcal{C} \right|}$ subsets of $\mathcal{C}$ that represents all possible combinations of candidates. 
\begin{equation}
    \begin{split}
        P \left( \mathcal{C} \right) = 2^\mathcal{C} = &\{ \left\{ v_1 \right\}, \left\{ v_2 \right\},\cdots, \left\{ v_{\left| \mathcal{C} \right|} \right\}, \\
        & \left\{ v_1, v_2 \right\}, \left\{ v_1, v_3 \right\},\cdots, \left\{ v_{\left| \mathcal{C} \right|-1}, v_{\left| \mathcal{C} \right|} \right\}, \\
        & \left\{ v_1, v_2, v_3 \right\} \cdots, \left\{ v_{\left| \mathcal{C} \right|-2}, v_{\left| \mathcal{C} \right|-1}, v_{\left| \mathcal{C} \right|} \right\}, \\
        & \cdots, \\
        & \left\{ v_1, \cdots, v_{\left| \mathcal{C} \right|-1} \right\}, \cdots, \left\{ v_2, \cdots, v_{\left| \mathcal{C} \right|} \right\}, \\
        & \mathcal{C} \}.
    \end{split}
\end{equation} 
Each belief $m_\Omega$ is exactly assigned to the proposition $\Omega$ and to no smaller subset. In this work, the beliefs are assigned to all the subsets of $\mathcal{C}$, whose sum equals 1, and no belief is left to the empty set. Notably, the belief exactly assigned to the entire set $\mathcal{C}$ is the degree of global ignorance, while belief assigned to a non-singleton subset of $\mathcal{C}$ (excluding $\mathcal{C}$ itself) indicates the degree of local ignorance. When both global and local ignorance are absent, the belief function degenerates into a classical probability distribution.

To identify the most likely source, we employ a generalized Bayesian inference model that integrates $2 \left | \mathcal{O}^+ \right |$ pieces of information collected from exposed observers, represented as $\mathbf{G} = \left\{ \mathbf{G}_{1}, \mathbf{G}_{2}, \cdots, \mathbf{G}_{l}, \cdots, \mathbf{G}_{2 \left | \mathcal{O}^+ \right |} \right\}$, each associated with a credibility score $\Psi = \left\{ \psi_1, \psi_2, \cdots, \psi_l, \cdots, \psi_{2 \left | \mathcal{O}^+ \right |} \right\}$. Because each observer occupies a different structural position in the network, the reliability of their reported data varies. To account for this, each likelihood input is weighted by a credibility coefficient derived from the observer's local topological characteristics, thereby reducing the influence of noisy or unreliable observations. Specifically, 
\begin{itemize}
    \item For pairwise likelihood information $\mathbf{G}_{\ell_i \ell_j}$ involving two observers, the credibility $\psi_{\ell_i \ell_j}$ is defined as
    \begin{equation}
        \psi_{\ell_i \ell_j} = \frac{k_{o_i} + k_{o_j}}{2 \times \max_{o_m \in \mathcal{O}^+} k_{o_m}},
    \end{equation}
    where $k_{o_i}$ denotes the number of one-hop neighbors of observer $o_i$. Observers with broader local connectivity, demonstrated by a higher degree, are considered to provide more reliable information. 
    \item For group likelihood information $\mathbf{G}_{\setminus \ell_i}$ derived from a group of observers $\mathcal{O}^+ \setminus o_i$, $\psi_{\setminus \ell_i}$ is reduced when the group lacks a highly connected (authoritative) observer,
    \begin{equation}
        \psi_{\setminus \ell_i} = \frac{1 - k_{o_i} / \sum_{o_m \in \mathcal{O}^+} k_{o_m}}{2}.
    \end{equation}
    This reflects the decreased reliability of likelihood estimates when a key observer is absent.
\end{itemize}

Each information $\mathbf{G}_{l}$ yields a weighted belief distribution,
\begin{equation}
    S_l = \{ \left( \Omega_l(\zeta_n), m_{\zeta_n, l} \right), n = 1, \cdots, N; \left( P \left( \mathcal{C} \right), m_{P \left( \mathcal{C} \right), l} \right) \},
\end{equation}
where $\Omega_l(\zeta_n) \subseteq \mathcal{C}$ represents a set of candidates supported by the belief function $m_{\zeta_n, l}$, satisfying $\sum_n m_{\zeta_n, l} + m_{P \left( \mathcal{C} \right), l} = 1$. Candidates with similar likelihoods (within a threshold $\phi < 1$) are grouped into the same cluster $\Omega_l(\zeta_n) = \{v \in \mathcal{C} : |\mathcal{P}_l(v) - \zeta_n| \leq \phi\}$. These clusters are mutually exclusive and collectively exhaustive, forming $N_l$ disjoint subsets $\Omega_l = \{\Omega_l(\zeta_1), \ldots, \Omega_l(\zeta_{N_l})\}$ such that $\Omega_l(\zeta_i) \cap \Omega_l(\zeta_j) = \emptyset$ for $\forall i \neq j$, and $\bigcup_i \Omega_l(\zeta_i) = \mathcal{C}$. The belief function supporting the proposition that $\Omega_l \left( \zeta_n \right)$ contains the true source is,
\begin{equation}
    m_{\zeta_n, l} = \psi_l \times \sum_{v \in \Omega_l \left( \zeta_n \right)} \frac{\mathcal{P}_l \left( v \right)}{\left| \Omega_l \left( \zeta_n \right) \right|},
\end{equation}
where $\psi_l$ denotes the credibility of the $l$-th likelihood $\mathbf{G}_{l}$. 

The generalized Bayesian inference model~\cite{yang2013evidential} recursively combines these weighted belief distributions,
\begin{equation}
    S_{e \left( 2 \left | \mathcal{O}^+ \right | \right)} = S_1 \oplus S_2 \oplus \cdots \oplus S_{2\left | \mathcal{O}^+ \right |}, 
\end{equation}
where $\oplus$ denotes the recursive combination operator, and $S_{e \left( 2 \left | \mathcal{O}^+ \right | \right)}$ represents the aggregated belief distribution. At iteration $l$, the combination between $S_{e \left( l - 1 \right)}$ (the result of the previous $l - 1$ combinations) and $S_l$ follows, 
\begin{equation}
    \begin{split}
        \hat{m}_{\theta, e(l)} &= [(1-\psi_l) m_{\theta, e(l-1)} + m_{P(\mathcal{C}),e(l-1)} m_{\theta, l}] \\
        &+ \sum_{B \cap C = \theta} m_{B,e(l-1)} m_{C,l}, \forall \theta \subseteq \mathcal{C}, \\
        \hat{m}_{P(\mathcal{C}), e(l)} &= (1 - \psi_l) m_{P(\mathcal{C}), e(l-1)},
    \end{split}
\end{equation}
and the result is normalized as
\begin{equation}
    m_{\theta, e(l)} = \frac{\hat{m}_{\theta, e(l)}}{\sum_\theta \hat{m}_{\theta, e(l)} + \hat{m}_{P(\mathcal{C}), e(l)}}.
\end{equation}
The process begins with $m_{\theta, e(1)} = m_{\theta, 1}$ and $m_{P(\mathcal{C}),e(1)} = m_{P(\mathcal{C}),1}$. Here, $\theta \subseteq \mathcal{C}$ serves as a general representation of $\Omega_l(\zeta_n) \subseteq \mathcal{C}$ for notational simplicity.

After all belief functions are combined, the residual belief $m_{P \left( \mathcal{C} \right), e \left( 2 \left | \mathcal{O}^+ \right | \right)}$ that represents uncertainty unassigned to any specific subset is removed before final decision-making,
\begin{equation}
    m_{\theta, e \left( 2 \left | \mathcal{O}^+ \right | \right)} = \frac{m_{\theta, e \left( 2 \left | \mathcal{O}^+ \right | \right)}}{1 - m_{P \left( \mathcal{C} \right), e \left( 2 \left | \mathcal{O}^+ \right | \right)}} = \frac{m_{\theta, e \left( 2 \left | \mathcal{O}^+ \right | \right)}}{\sum_{\theta} m_{\theta, e \left( 2 \left | \mathcal{O}^+ \right | \right)}}.
\end{equation}
The candidate $v_i$ with the highest combined belief,
\begin{equation}
    \eta_{v_i} = p_{v_i, e \left( 2 \left | \mathcal{O}^+ \right | \right)} = \sum_{v_i \in \theta, \forall \theta} \frac{m_{\theta, e \left( 2 \left | \mathcal{O}^+ \right | \right)}}{\left| \theta \right|},
\end{equation}
is then identified as the most likely source $\hat{s}$. This inference process effectively aggregates multi-observer evidence while accounting for both credibility and uncertainty, enabling robust and interpretable source localization. Details of the generalized Bayesian inference procedure can be found in~\cite{yang2013evidential}.

From the set of exposed observers, a total $2 \left | \mathcal{O}^+ \right | \ll \left | \mathcal{N} \right |$ belief functions are derived from pairwise and group likelihoods. Each belief function is obtained via a clustering algorithm with computational complexity $O \left( \left | \mathcal{C} \right | \log \left | \mathcal{C} \right | \right)$ for classifying candidates $\mathcal{C}$, resulting in a total generation cost of $O \left( \left | \mathcal{O}^+ \right | \left | \mathcal{C} \right | \log \left | \mathcal{C} \right | \right)$. Given that each belief function contains $N_l$ focal elements, the integration step incurs an additional $O \left( \left | \mathcal{O}^+ \right | N_l^2 \right)$ cost. Therefore, the overall computational complexity is $O \left( \left | \mathcal{O}^+ \right | \left( \left | \mathcal{C} \right | \log \left | \mathcal{C} \right | + N_l^2 \right) \right)$.

To further improve scalability in large networks, subsets with negligible belief are pruned to focus computational resources on the most plausible candidates. As in many complex systems, the distribution of belief function exhibits a ``rich-get-richer'' pattern, where only a few subsets accumulate high belief, while most carry negligible support. Hence, when the belief set becomes excessively large, subsets with the lowest cumulative belief collectively accounting for a small threshold $\upsilon = 2\% \ll 1$ are iteratively removed, streamlining computation while preserving inference accuracy.

\section{Experimental Analysis} \label{sec.experiment}


\subsection{Illustrations of Data, Comparative Methods, and Propagation Patterns}

To evaluate the applicability and effectiveness of the developed EAGLE framework, we employ three synthetic networks and ten real-world networks spanning diverse domains and structural characteristics \footnote{\url{https://networkrepository.com}}$^,$ \footnote{\url{http://snap.stanford.edu/data}}. 
In addition, the large-scale Twitter datasets capturing real information propagation processes are incorporated~\cite{omodei2015characterizing}. These correspond to the MoscowAthletics dataset, which records the discussion during the 14th IAAF World Championships in Athletics held in Moscow, and NYClimateMarch dataset, documenting online activism related to global climate action in New York. In both cases, only forwarding interactions are considered, as they are the dominant mechanism through which information propagates to broader audiences. When a specific user is selected as the source, information is allowed to spread to this user's one-hop neighbors if a forwarding event between them is observed at a particular timestamp. To avoid redundant propagation loops, only the first instance of a user forwarding another's post is retained; subsequent forwards between the same user pair are ignored. This process, where exposed users forward the information to their unexposed neighbors, continues iteratively until no further users can be reached. In this way, we reconstruct the real propagation paths initiated by individual users on Twitter. 

For all datasets, only the largest connected component without self-loops is analyzed. The key statistical characteristics of these networks, including the number of nodes ($\left | \mathcal{N} \right |$), number of edges ($\left | \mathcal{E} \right |$), network density ($\rho$), average degree ($\bar{k}$), average clustering coefficient ($\bar{C}$), and average shortest path length ($\bar{d}$), are summarized in Table \ref{tab:StatisticalNetwork}.

\begin{table}[!tb]
    \centering
    \caption{Statistical characteristics of social networks.}
    \label{tab:StatisticalNetwork}
    \begin{tabular}{ccccccc}
        \hline
        Network & $\left | \mathcal{N} \right |$ & $\left | \mathcal{E} \right |$ & $\rho$ & $\bar{k}$ & $\bar{C}$ & $\bar{d}$ \\
        \hline
        dolphins & 62 & 159 & 0.0841 & 5.1290 & 0.2590 & 3.3570 \\
        USAir97 & 332 & 2126 & 0.0387 & 12.8072 & 0.6252 & 2.7381 \\
        email-Eu & 986 & 16,064 & 0.0331 & 32.5842 & 0.4071 & 2.5869 \\
        polblogs & 1,222 & 16,714 & 0.0224 & 27.3552 & 0.3203 & 2.7375 \\
        facebook & 4,039 & 88,234 & 0.0108 & 43.6910 & 0.6055 & 3.6925 \\
        ca-GrQc & 4,158 & 13,422 & 0.0016 & 6.4560 & 0.5569 & 6.0494 \\
        wiki-Vote & 7,066 & 100,736 & 0.0040 & 28.5129 & 0.1419 & 3.2475 \\
        lastfm-asia & 7,624 & 27,806 & 0.0010 & 7.2943 & 0.2194 & 5.2322 \\
        ca-HepTh & 8,638 & 24,806 & 0.0007 & 5.7435 & 0.4816 & 5.9454 \\
        ca-HepPh & 11,204 & 117,619 & 0.0019 & 20.9960 & 0.6216 & 4.6727 \\
        MoscowAthletics & 66,474 & 98,959 & $4.48 \times 10^{-5}$ & 2.9774 & 0.0690 & 4.7183 \\
        NYClimateMarch & 92,139 & 211,256 & $4.98 \times 10^{-5}$ & 4.5856 & 0.0723 & 4.0495 \\
        \hline
    \end{tabular}
\end{table}

To comprehensively benchmark EAGLE, we compare its performance against several SOTA observer-based source localization methods widely recognized in the literature. Specifically, Sequential Neighbor Filtering (SNF)~\cite{wang2020locating} identifies the source by measuring the Pearson correlation coefficient between the minimum informed time vector and node radii once all observers have been exposed. Ordinary Least Squares (OLS)~\cite{lalou2016least} assumes a linear relationship between infection times and effective distances to infer the source, requiring the knowledge of the diffusion start time, which is an often impractical assumption. Gaussian Estimator (GE)~\cite{pinto2012locating} utilizes maximum likelihood estimation under both observed and deterministic delays, which is ideally suited for tree-like structures under the heterogeneous diffusion process with known means and variances, with limited applicability to real-world networks. Direction-induced Search based Gaussian Estimator (DISGE)~\cite{yang2020locating} is an extension of GE that integrates directional information, inheriting similar structural limitations. Evidential Source Localization (ESL)~\cite{zhao2023early} aggregates observer data via a maximum likelihood-based evidential approach to consider the involved uncertainty and noise, though at the cost of substantial computational complexity. These methodologies represent the current methodological frontier for source localization and collectively provide robust benchmarks for assessing EAGLE's performance across varying network topologies and propagation dynamics.

The robustness of EAGLE is evaluated under four representative propagation models~\cite{wang2022rapid,pinto2012locating,wang2020locating,ji2023signal}, each capturing distinct diffusion mechanisms, including Heterogeneous SI Epidemic Model (HSI, edge infection rate follow a uniform distribution $U(0.1, 0.15)$), Heterogeneous Diffusion Process (HDP, edge propagation delays are drawn from a normal distribution $\mathcal{N}(4,1^2)$), Random Walk Propagation Process (RWP, using $\left| \mathcal{N} \right| / 2$ random walkers), and Time-varying Network Diffusion Process (TDP, edge activation rate and time delays are drawn from $U(0.1, 0.2)$ and $U(1,4)$, respectively). These models collectively enable a rigorous evaluation of EAGLE's adaptability and robustness across a wide range of scenarios, from epidemic spread to information diffusion, capturing the temporal and structural complexities inherent in large-scale social systems. The validity and representativeness of these models have been extensively verified in prior studies~\cite{wang2020locating,ji2023signal}. 

All experiments are conducted on a laptop equipped with 11th Gen Intel(R) Core(TM) i7-1165G7 @2.80GHz and 16GB RAM. The algorithms are implemented in Python, leveraging NetworkX 2.8.4 and igraph 0.10.2 for network representation and computation.

\subsection{Early Detection of Propagation via Effective Observer Deployment}

Effective observer deployment is crucial for the timely detection and tracing of misinformation sources. To evaluate the efficiency of our deployment strategy, we measure how quickly observers can detect ongoing information spreading across ten real-world networks. 
The proposed strategy is compared against a genetic algorithm (GA)-based optimization approach with the same fitness function and two traditional strategies, including random and degree-based deployment, under four propagation models: HSI, HDP, RWP, and TDP. 

\begin{figure}[!tb]
\begin{center}
\includegraphics[width=\textwidth]{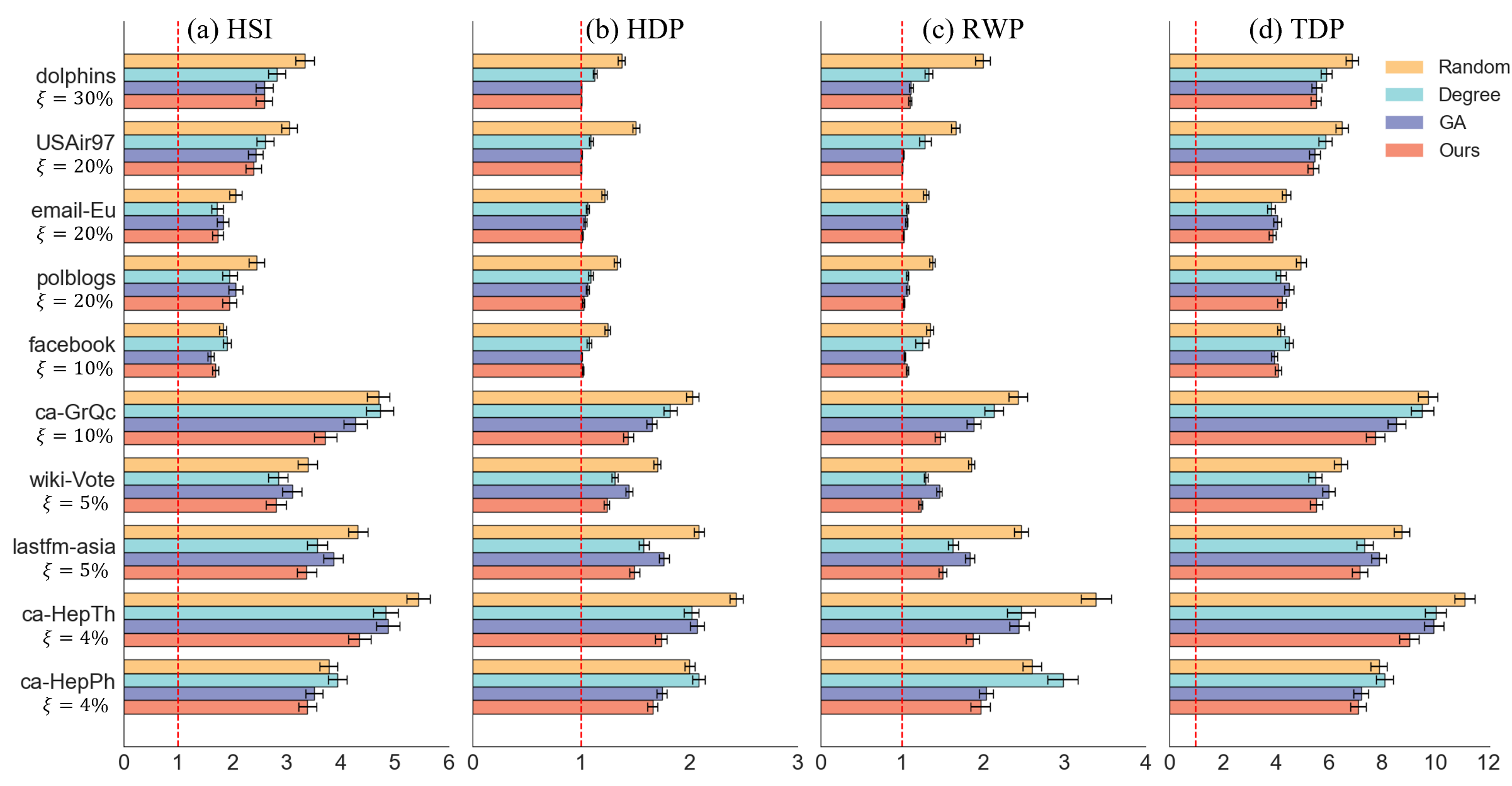}
\end{center}
\caption{\textbf{Effectiveness of observer deployment strategies across real-world networks.} The average time step of detection over 1000 realizations shows comparisons of four deployment strategies under four distinct propagation patterns. In each realization, a random propagation source, excluded from any observer set, is selected.} 
\label{fig:TimeStamp}
\end{figure}

The results are illustrated in \fref{fig:TimeStamp}, where the detection timestamp is calculated as $\min t_i - t_s$ for $\forall o_i \in \mathcal{O}$. Note that the propagation start time $t_s$ is known here only for benchmarking deployment performance in \fref{fig:TimeStamp}; in all subsequent analysis, $t_s$ remains unknown, reflecting realistic operational conditions. The red dashed line indicates the ideal scenario: immediate observation upon propagation onset ($\min t_i - t_s = 1$). The ratio of observers $\xi$, also shown in \fref{fig:TimeStamp}, is kept small in large networks to minimize costs but increased in smaller networks to ensure adequate coverage, allowing operators to flexibly balance between efficiency and budget constraints. 

Across all networks and propagation patterns, random deployment exhibits the slowest detection performance, followed by the degree-based strategy. Both conventional approaches lag significantly behind optimization-based methods. Compared with the GA strategy, our proposed deployment achieves consistently shorter detection times in nearly all cases, with the advantage particularly pronounced in large networks. Although the GA approach occasionally performs comparably, such as in the facebook network, it often exhibits instability, even underperforming the degree-based approach in networks such as email-Eu, wiki-Vote, and lastfm-asia. In contrast, our method delivers the most stable and superior detection performance across almost all scenarios, demonstrating both robustness and scalability. Notably, the performance gap between methods widens in large networks, while in smaller ones the differences are less pronounced, highlighting the growing importance of optimized deployment as network complexity increases. Moreover, under the HDP and RWP models, optimized deployment in the smaller networks can detect propagation immediately upon onset (\fref{fig:TimeStamp}(b) and (c)), achieving the theoretical optimum. This indicates that the optimization reliably positions at least one observer within one hop of any potential source, ensuring rapid detection. Overall, these findings confirm that our deployment strategy consistently outperforms all baselines across diverse networks and propagation models.


\begin{figure}[!tb]
\begin{center}
\includegraphics[width=0.6\textwidth]{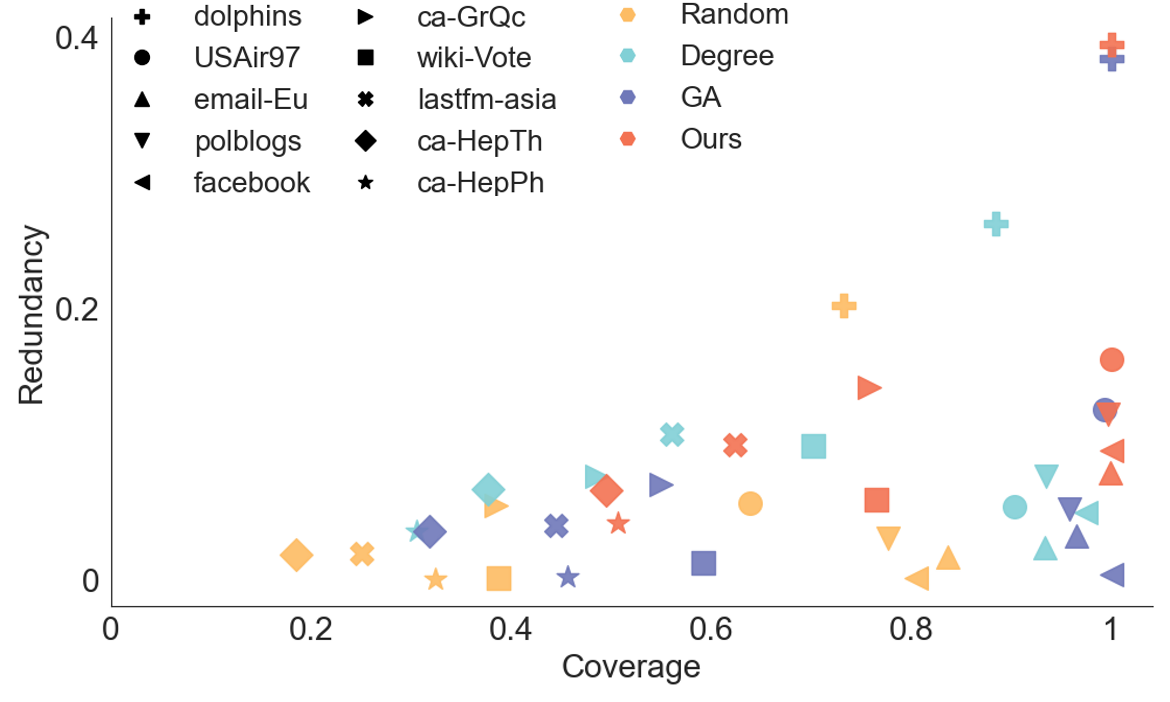}
\end{center}
\caption{Trade-off between coverage and redundancy in observer deployment strategies across real-world networks.}
\label{fig:TradeOff}
\end{figure}


To further assess the quality of the observer deployment, we analyze the coverage-redundancy trade-off of each deployment strategy in \fref{fig:TradeOff}. The objective is twofold: 1) maximize coverage, that is, the fraction of candidate nodes with at least one observer among their one-hop neighbors; and 2) maintain adequate redundancy, ensuring multiple observers can cross-monitor the same candidate, quantified as $F \left( \mathcal{O} \right) / \left( \left | \mathcal{N} \right | - \left | \mathcal{O} \right | \right)$ with $\lambda=0$.
As shown in \fref{fig:TradeOff}, the proposed strategy (highlighted in red) achieves the highest coverage and competitive redundancy across all networks. In the smaller networks, it attains nearly complete coverage (approaching one) and high redundancy, while in the larger networks it substantially surpasses all comparison methods in coverage. Although its redundancy is marginally lower than that of the degree-based strategy in the wiki-Vote, lastfm-asia, and ca-HepTh networks, this is expected because degree-based deployment inherently clusters observers around hubs, inflating redundancy but limiting global reach. In contrast, our approach yields a more balanced and strategic observer distribution, maximizing coverage while maintaining sufficient redundancy to ensure robust cross-validation of observations. These results demonstrate that the optimization-driven deployment effectively allocates observers across the network to achieve comprehensive coverage with sufficient redundancy. Hence, this strategic deployment mechanism forms the foundation of early detection and rapid response to misinformation propagation in complex systems.

\subsection{Inference of Propagation Sources}

We evaluate the performance of the developed EAGLE framework in inferring propagation sources by comparing it with SOTA methods using two metrics. First, we measure the average error distance between the real source $s^*$ and the set of inferred potential sources $\hat{\bm{s}} = \left\{ \hat{s}_1, \hat{s}_2, \cdots, \hat{s}_n \right\}$, defined as $\bar{d} = \sum_{s \in \hat{\bm{s}}} d\left( s, s^* \right)/\left| \hat{\bm{s}} \right|$, where $d\left( s, s^* \right)$ denotes the shortest path length between two users. Second, we employ the F-score, which harmonizes Precision and Recall into a balanced metric, $\text{F-score} = \left( 1 + \beta \right)^2 \times \left(\text{Precision} \times \text{Recall}\right)/\left(\beta^2 \times \text{Precision} + \text{Recall}\right)$, where $\beta=1$ indicates equal weighting. Here, Precision quantifies the accuracy within the inferred source set, while Recall measures the fraction of true sources correctly identified. When the propagation source is perfectly inferred, meaning only one candidate is inferred and it matches the true source, the error distance equals zero and the F-score equals one.

\begin{table*}[!tb]
\centering
\caption{Error distance of different approaches on synthetic networks under varying propagation patterns.}
\resizebox{\textwidth}{!}{
\begin{tabular}{ccccccccc|ccccccccc}
\hline
Spread & Network & $\rho$ & SNF & OLS & GE & DISGE & ESL & EAGLE & Spread & Network & $\rho$ & SNF & OLS & GE & DISGE & ESL & EAGLE \\ \hline
\multirow{9}{*}{HSI} & \multirow{3}{*}{BA} & 0.002 & 3.24 & 2.81 & 1.83 & \underline{1.69} & 2.97 & \textbf{0.57} & \multirow{9}{*}{HDP} & \multirow{3}{*}{BA} & 0.002 & 3.11 & 0.28 & 0.23 & \textbf{0.04} & 0.18 & \underline{0.13} \\ 
 &  & 0.005 & 2.75 & 2.46 & 1.68 & \underline{1.62} & 2.03 & \textbf{0.81} &  &  & 0.005 & 2.73 & 0.13 & 0.14 & \textbf{0.00} & \textbf{0.00} & 0.62 \\ 
 &  & 0.01 & 2.68 & 1.78 & 1.94 & \underline{1.27} & 2.11 & \textbf{0.52} &  &  & 0.01 & 2.73 & \underline{0.01} & 0.20 & \textbf{0.00} & 0.06 & 1.18 \\ \cline{2-9} \cline{11-18} 
 & \multirow{3}{*}{ER} & 0.002 & 4.05 & 3.96 & \underline{1.72} & 2.38 & 3.02 & \textbf{1.56} &  & \multirow{3}{*}{ER} & 0.002 & 4.39 & \textbf{0.05} & 0.20 & 0.39 & 1.08 & \underline{0.08} \\ 
 &  & 0.005 & 2.81 & 3.13 & \underline{1.74} & 2.20 & 2.63 & \textbf{1.17} &  &  & 0.005 & 3.13 & 0.05 & 0.03 & \textbf{0.00} & \textbf{0.00} & 0.47 \\ 
 &  & 0.01 & 2.20 & 2.86 & 1.70 & \underline{1.69} & 2.11 & \textbf{0.75} &  &  & 0.01 & 2.79 & 0.14 & \textbf{0.00} & \textbf{0.00} & \textbf{0.00} & 1.55 \\ \cline{2-9} \cline{11-18} 
 & \multirow{3}{*}{WS} & 0.002 & 3.76 & 3.72 & \underline{1.35} & 1.87 & 3.06 & \textbf{1.32} &  & \multirow{3}{*}{WS} & 0.002 & 2.23 & \textbf{0.01} & 0.56 & 0.90 & 2.03 & \underline{0.02} \\ 
 &  & 0.005 & 2.63 & 2.75 & \underline{1.15} & 1.32 & 1.76 & \textbf{0.81} &  &  & 0.005 & 3.45 & \textbf{0.02} & 0.11 & 0.13 & 0.06 & \textbf{0.02} \\
 &  & 0.01 & 1.79 & 2.24 & 1.28 & \underline{0.87} & 1.35 & \textbf{0.53} &  &  & 0.01 & 2.02 & 0.04 & \underline{0.02} & \textbf{0.00} & 0.04 & 0.22 \\ \hline
\multirow{9}{*}{RWP} & \multirow{3}{*}{BA} & 0.002 & \underline{0.56} & 2.48 & 1.13 & 0.80 & 1.68 & \textbf{0.20} & \multirow{9}{*}{TDP} & \multirow{3}{*}{BA} & 0.002 & 3.19 & 2.33 & 1.91 & \underline{1.83} & 2.92 & \textbf{0.77} \\ 
 &  & 0.005 & \underline{0.26} & 2.40 & 1.45 & 1.09 & 1.19 & \textbf{0.05} &  &  & 0.005 & 2.88 & 2.70 & 2.02 & \underline{1.99} & 2.20 & \textbf{0.75} \\ 
 &  & 0.01 & \underline{1.53} & 2.17 & 1.81 & 1.89 & 1.66 & \textbf{0.13} &  &  & 0.01 & 2.51 & 2.22 & 1.97 & \underline{1.79} & 2.23 & \textbf{0.70} \\ \cline{2-9} \cline{11-18} 
 & \multirow{3}{*}{ER} & 0.002 & 1.14 & 4.44 & \textbf{0.92} & 1.65 & 2.28 & \underline{0.99} &  & \multirow{3}{*}{ER} & 0.002 & 3.91 & 3.67 & \textbf{1.35} & 2.24 & 2.93 & \underline{1.62} \\ 
 &  & 0.005 & \underline{0.77} & 3.21 & 0.84 & 1.51 & 1.94 & \textbf{0.29} &  &  & 0.005 & 2.98 & 3.27 & \underline{1.89} & 2.31 & 2.89 & \textbf{1.62} \\ 
 &  & 0.01 & \underline{0.51} & 2.84 & 1.09 & 1.49 & 2.15 & \textbf{0.39} &  &  & 0.01 & 2.50 & 2.89 & 2.11 & \underline{2.06} & 2.47 & \textbf{1.66} \\ \cline{2-9} \cline{11-18} 
 & \multirow{3}{*}{WS} & 0.002 & 1.47 & 5.46 & \textbf{1.11} & 1.65 & 2.39 & \underline{1.34} &  & \multirow{3}{*}{WS} & 0.002 & 3.25 & 1.97 & \underline{1.40} & 1.75 & 2.39 & \textbf{0.89} \\ 
 &  & 0.005 & 1.18 & 3.52 & \underline{0.95} & 1.09 & 1.62 & \textbf{0.84} &  &  & 0.005 & 2.85 & 2.82 & \underline{1.23} & 1.51 & 1.99 & \textbf{1.10} \\ 
 &  & 0.01 & \textbf{0.82} & 2.75 & 1.18 & 1.42 & 1.59 & \underline{1.05} &  &  & 0.01 & 2.28 & 2.39 & 1.45 & 1.45 & \underline{1.38} & \textbf{1.14} \\ \hline
\end{tabular}}
\label{tab:errordis}
\end{table*}

We first examine EAGLE's performance in synthetic networks with adjustable connectivity, including Barab{\'a}si--Albert (BA) scale-free, Erd{\H{o}}s--R{\'e}nyi (ER) random, and Watts–Strogatz (WS) small-world networks. The structural characteristics of these typical networks are detailed in~\cite{newman2018networks}. All networks are equipped with identical parameters ($\left | \mathcal{N} \right | = 2,000, \xi = 15\%, \varphi = 10 \%$) while varying the network density $\rho$ to manipulate connectivity. Each reported result is averaged over 100 realizations. The outcomes are summarized in~\tref{tab:errordis}, where the best performing method is highlighted in \textbf{bold}, and the second-best \underline{underlined}. As connectivity increases, EAGLE shows no uniform trend in error distance: its accuracy remains stable or fluctuates under most propagation patterns, with the exception of the HDP model, where its accuracy slightly decreases and is lower than the comparison approaches. This is expected, as methods such as GE and DISGE are specifically designed for HDP by assuming known propagation delay distributions (mean and covariance). Despite this, EAGLE achieves the lowest error distance in the majority of patterns and network types, highlighting its broad adaptability and robust capability across diverse network topologies. 

\begin{figure}[!tb]
\begin{center}
\includegraphics[width=\textwidth]{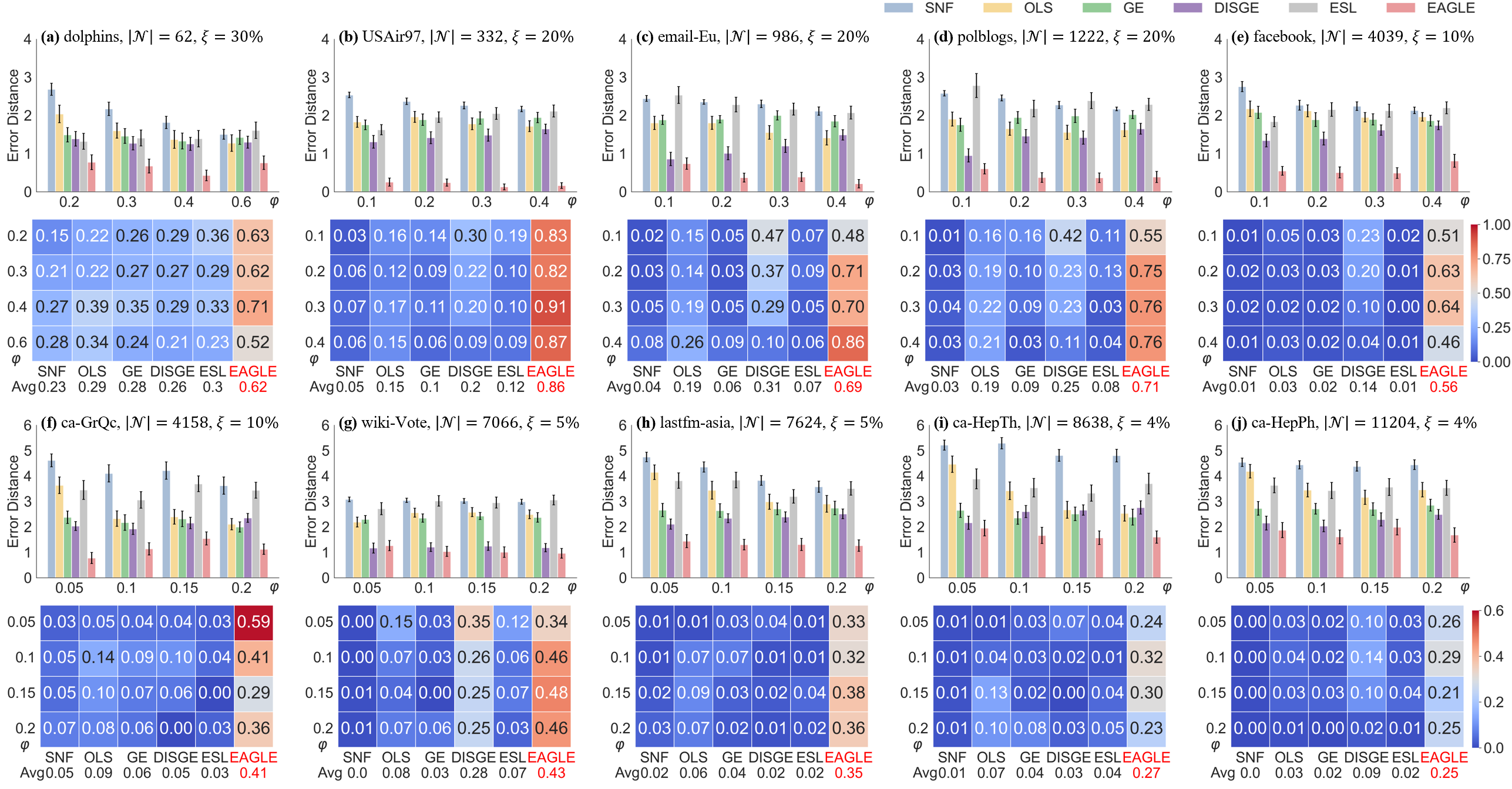}
\end{center}
\caption{\textbf{Comparative performance of EAGLE and benchmark approaches across networks of varying proportions of exposed observer under the HSI model.} The performance is evaluated by error distance (bar chart) and F-score (heat map) across ten real-world networks over 100 realizations: \textbf{(a)} dolphins, \textbf{(b)} USAir97, \textbf{(c)} email-Eu, \textbf{(d)} polblogs, \textbf{(e)} facebook, \textbf{(f)} ca-GrQc, \textbf{(g)} wiki-Vote, \textbf{(h)} lastfm-asia, \textbf{(i)} ca-HepTh, and \textbf{(j)} ca-HepPh. Error bars represent $95\%$ confidence interval.}
\label{fig:Localization}
\end{figure}

We further compare EAGLE against the SOTA methods under the HSI model and varying ratios of exposed observers $\varphi$, and results are shown in \fref{fig:Localization}. In smaller networks, EAGLE consistently outperforms all benchmarks, maintaining an average error distance below 1 and an F-score exceeding 0.5, while competing methods rarely achieve F-scores above 0.5. As network size increases, the performance of competing methods such as SNF, GE, and ESL diminishes significantly. For example, in the email-Eu network (fewer than 1,000 nodes), their F-scores fall below 0.1, whereas EAGLE achieves an F-score of 0.86, demonstrating exceptional accuracy (\fref{fig:Localization}(c)). For larger networks such as ca-GrQc (around 4,000 nodes), OLS and DISGE fail to produce precise source estimates, yielding error distances greater than 2 and F-scores below 0.1 (\fref{fig:Localization}(f)). 
Even in very large networks (over 4,000 to 10,000 nodes), including lastfm-asia, ca-HepTh, and ca-HepPh networks (\fref{fig:Localization}(h) -- (j)), EAGLE consistently achieves the lowest error distance (below 2) and highest F-score, often two to five times greater than those of competing approaches. The results demonstrate EAGLE's scalability and superior inference precision across a wide range of networks. 

\begin{figure*}[!tb]
\begin{center}
\includegraphics[width=\textwidth]{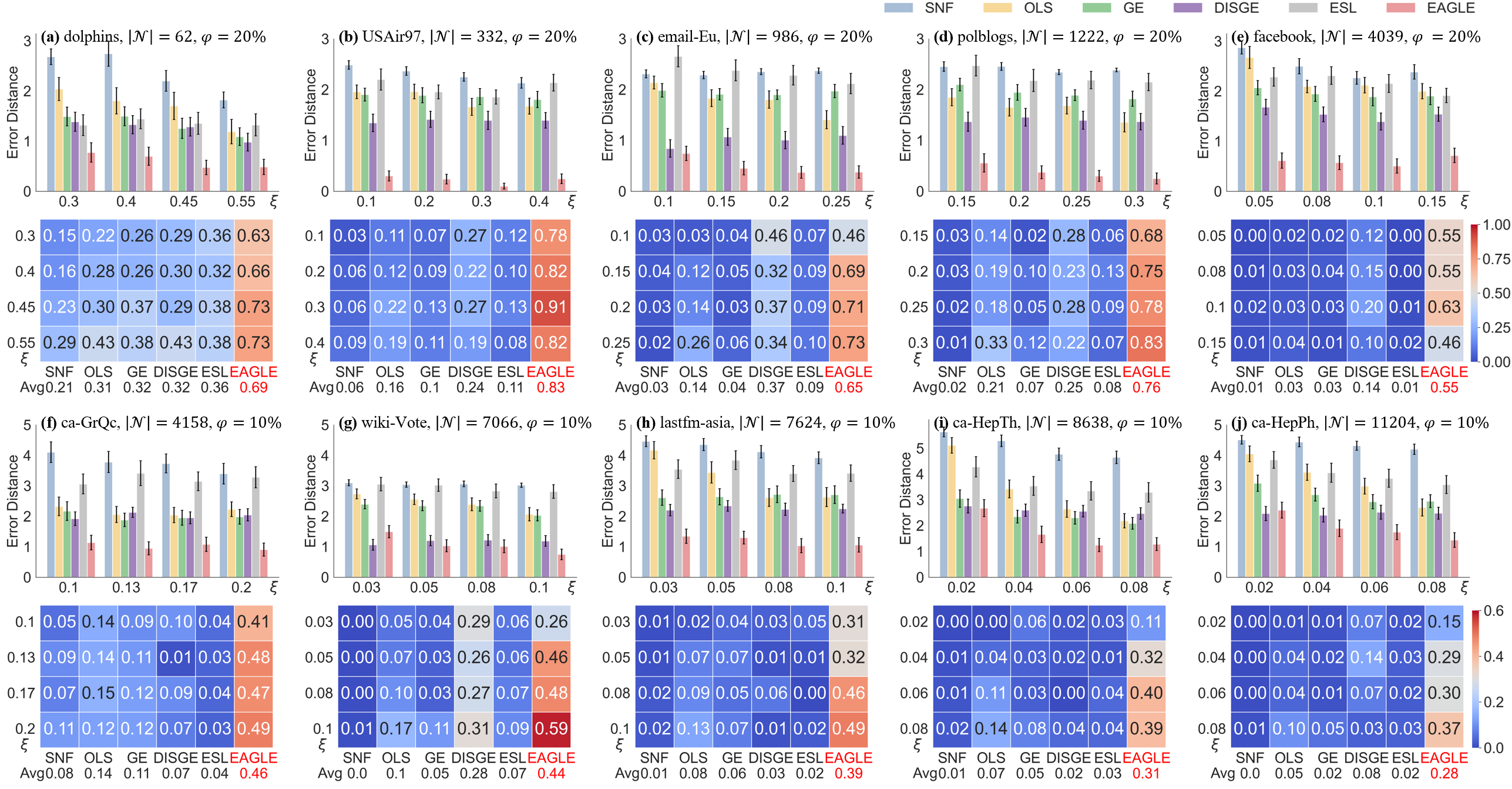}
\end{center}
\caption{Comparative performance of EAGLE and benchmark methods on networks under the HSI model with varying observer proportions.}
\label{fig:LocalizationHSI2}
\end{figure*}

The performance of the proposed approach is further validated under different experimental settings, demonstrating its applicability and robustness. Here, we reverse this design by fixing the ratio of exposed observers and adjusting the ratio of observers in \fref{fig:LocalizationHSI2}, thereby capturing how observation capacity influences inference accuracy. Across all scenarios, EAGLE achieves the best performance, consistently surpassing competing approaches. As the proportion of exposed observers decreases, error distance grows and F-score declines, reflecting the inherent challenge of inference under limited observation. This degradation is observed across all methods, highlighting a fundamental difficulty in source localization as the number of exposed observers reduces. In small networks with no more than 1,000 nodes, EAGLE produces significantly lower error distances than alternative methods, and its F-score is two to three times higher than the second-best approach. Even in networks of around 10,000 nodes, where the error distances of some methods approach those of EAGLE, their F-scores remain four to ten times lower, indicating that they can only approximate source locations, while EAGLE continues to provide more precise inference. These results highlight the robustness of EAGLE under varying observation capacities.

\begin{figure*}[!tb]
\begin{center}
\includegraphics[width=\textwidth]{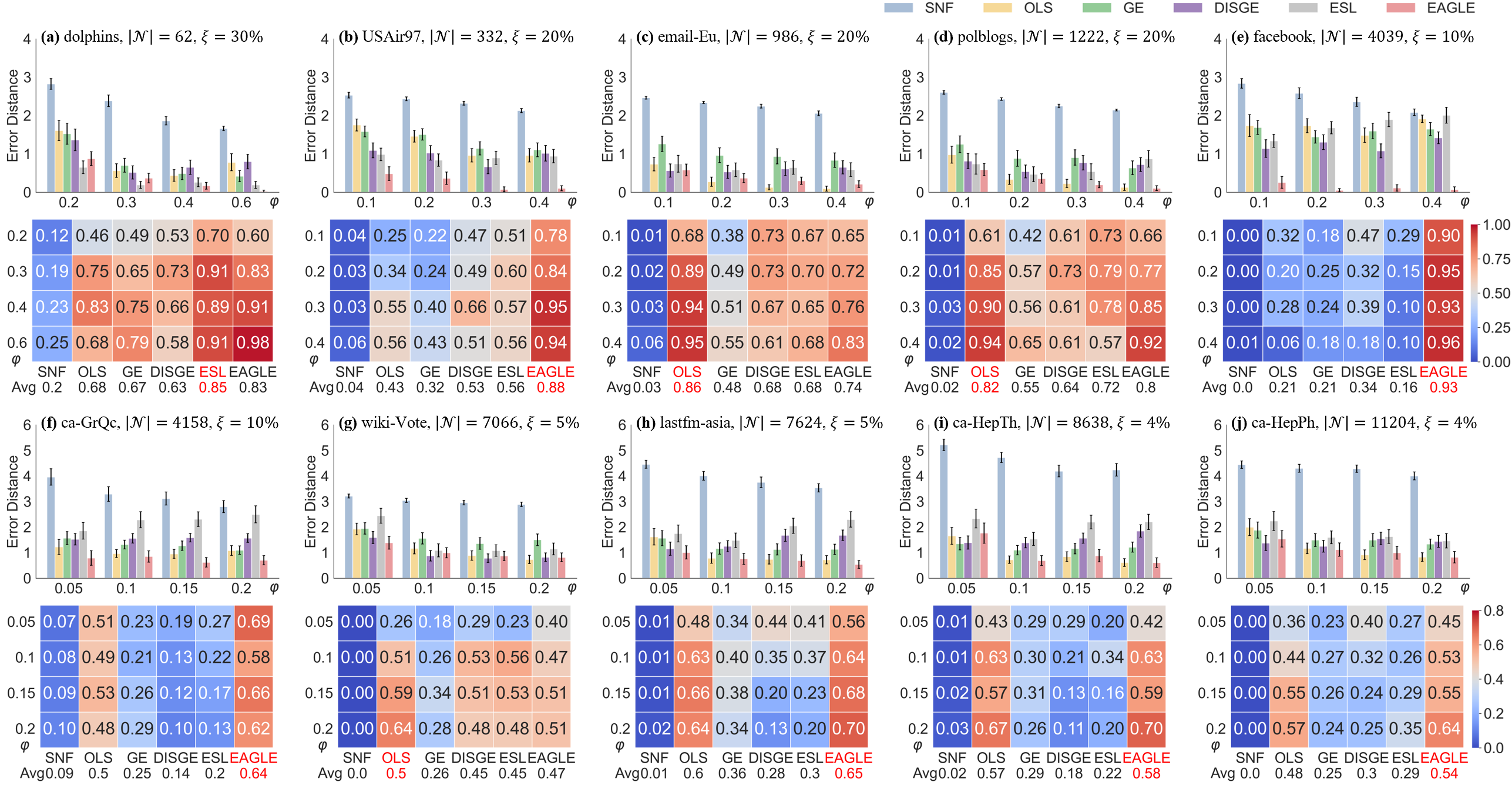}
\end{center}
\caption{Comparative performance of EAGLE and benchmark methods on networks under the HDP with varying exposed observer proportions.}
\label{fig:LocalizationHDP}
\end{figure*}

We further assess the performance of EAGLE under the heterogeneous diffusion process, an alternative propagation pattern, with varying proportions of exposed observers in \fref{fig:LocalizationHDP}. Results show that ESL achieves the lowest error distance and highest F-score in the smallest network (\fref{fig:LocalizationHDP}(a)), and OLS performs best in some medium-sized networks, including email-Eu, polblogs, and wiki-Vote (\fref{fig:LocalizationHDP}(c), (d), and (g)). In these cases, the performance of EAGLE is slightly behind the top method but still substantially stronger than the remaining approaches. In larger networks (\fref{fig:LocalizationHDP}(i) and (j)), EAGLE clearly outperforms all comparison methods, offering both higher accuracy and greater stability. These results indicate that while the relative performance of methods may fluctuate depending on network size and propagation dynamics, EAGLE consistently delivers strong and reliable performance across diverse scenarios, reinforcing its suitability for large-scale social networks.

Overall, other methods exhibited unstable or scenario-dependent behavior, but EAGLE maintained remarkably stable and accurate inference performance across all tested configurations. The results confirm that EAGLE delivers robust, scalable, and accurate propagation source localization across diverse network types and diffusion dynamics. Its ability to maintain high precision under uncertainty and limited observation conditions highlights its strong potential for real-world applications in large-scale social and information networks.

\subsection{Source Inference in Twitter with Real Propagation Paths}

Simulations of information diffusion often fail to fully capture the complexity, heterogeneity, and uncertainty of real-world propagation dynamics in social networks. Twitter, as one of the world's most active social platforms, provides a unique opportunity to validate source inference methods using actual propagation sources and paths derived from event-specific discussions. In this study, we analyze two large-scale Twitter networks on distinct topics, including sports and climate change, reflecting diverse communication patterns and engagement dynamics. These networks, consisting of 66,474 and 92,139 users respectively, represent the largest known real-world social networks with reconstructed propagation paths used for validating source localization methods.

\begin{figure}[!tb]
\begin{center}
\includegraphics[width=\textwidth]{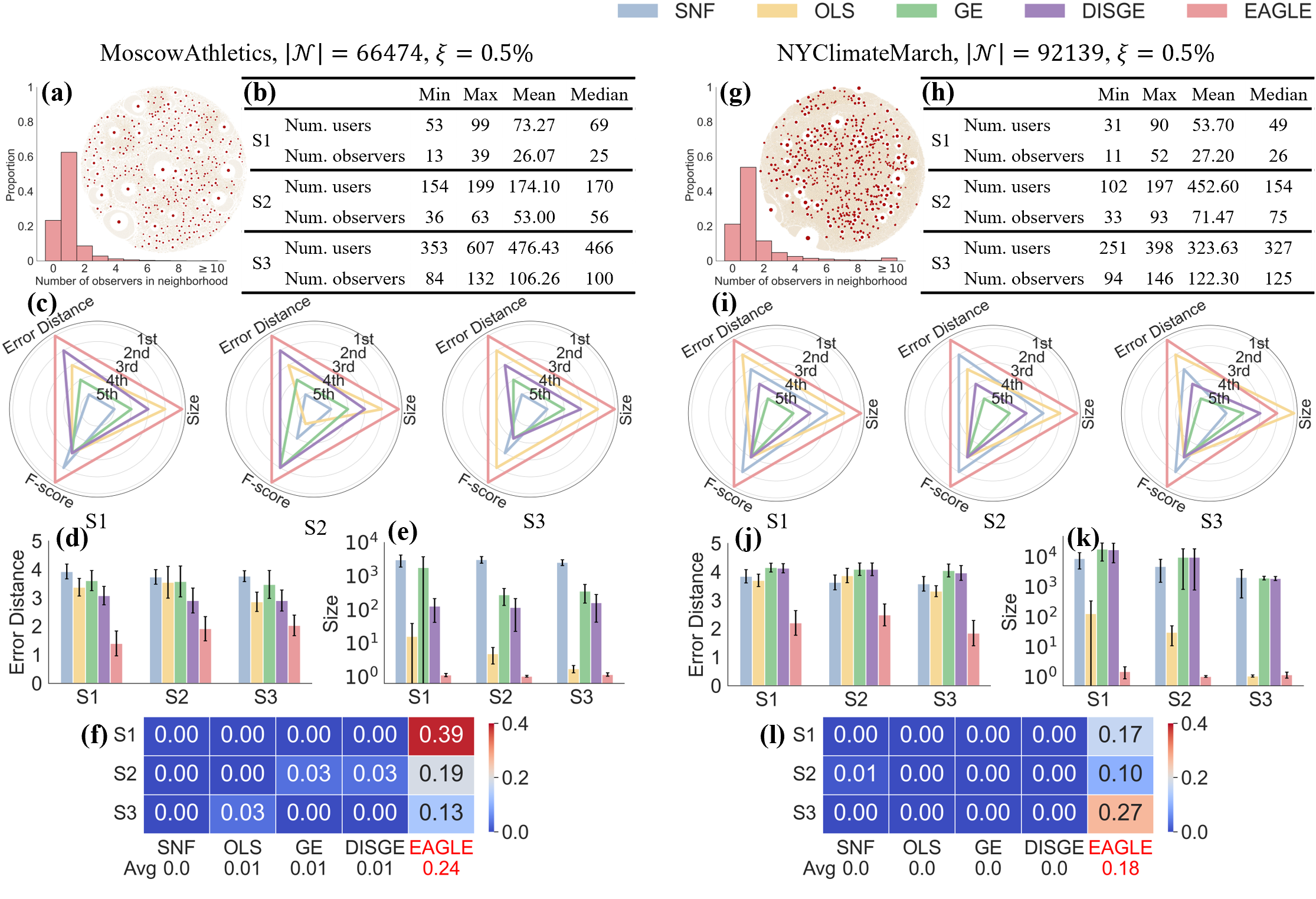}
\end{center}
\caption{\textbf{Evaluation of EAGLE and competing approaches using real propagation paths on Twitter.} Two real-world information propagation events serve as case studies to evaluate the practical effectiveness of EAGLE: \textbf{(a, g)} show the optimized deployment of observers (in red) and their one-hop observation coverage. \textbf{(b, h)} outline three experimental scenarios with varying numbers of users and observers, each repeated over 30 realizations, to assess the robustness under varying scales. \textbf{(c, i)} rank all approaches across three evaluative aspects, including \textbf{(d, j)} error distance, \textbf{(e, k)} size of inferred sources, and \textbf{(f, l)} F-score. Error bars denote $95\%$ confidence interval.} 
\label{fig:RealPath}
\end{figure}

The observers deployed by the developed optimization approach, using a $\xi = 0.5 \%$ observer ratio, are visualized in \fref{fig:RealPath}(a) and (g). Users with extensive connections are selected as observers, and they are strategically placed across the entire network rather than being concentrated in a small range, ensuring a broad observational scope. It is evident from the distribution of observer counts within one-hop neighbors, 
where only approximately $20\%$ of users have no observers among their one-hop neighbors, over $50\%$ have one observer, and the remaining users have multiple observers facilitating cross-observation of behavior.

These deployed observers are then utilized to infer the propagation source in these large-scale networks, each comprising over 60,000 users. To assess the efficacy of various localization strategies, three scenarios were constructed with an increasing number of exposed users and observers in the propagation, as depicted in \fref{fig:RealPath}(b) and (h). 
Source localization is attempted at the early stages of propagation when only a limited number of observers are exposed to the propagated information. It can be found that there are only 20 to 120 exposed observers in these scenarios, which consumes very little deployment resources to observe a wide range of user behaviors in networks. Due to the exponentially increasing computational demands of the ESL approach in large networks~\cite{zhao2023early}, it was excluded from these tests. The performance of the remaining approaches was evaluated based on the error distance, F-score, and the size of the inferred potential source sets, with rankings displayed in \fref{fig:RealPath}(c) and (i). EAGLE ranks first in almost all metrics and scenarios, and only ranks second in the size of potential sources in the third scenario (\fref{fig:RealPath}(i)). In contrast, the other methods exhibited varying performance across different scenarios, highlighting their inconsistency. Hence, the stable and superior performance of the developed EAGLE in inferring the propagation source is demonstrated in the comparison.

The performance of these approaches in two networks with real propagation paths is further discussed in three aspects. For the MoscowAthletics network, EAGLE reduced the error distance to half that of other methods in certain scenarios, consistently achieving smaller error distances across all scenarios as shown in \fref{fig:RealPath}(d). Moreover, the error distances obtained by EAGLE are consistently below 2, indicating close proximity to the real source, in contrast to other methods, which always exceed an error distance of 3. The difference in the size of the inferred potential source sets is also notable, as illustrated in \fref{fig:RealPath}(e). EAGLE can often pinpoint a unique potential source, thereby distinguishing the likelihood of users with close traits, while methods like SNF, GE, and DISGE identify over 1,000 possible candidates, which are inefficient for source localization. 
Despite some methods occasionally approaching the performance of EAGLE in terms of error distance or the size of potential sources, such as the size identified by OLS, their accuracy remains significantly lower (\fref{fig:RealPath}(f)). This trend is consistent in the NYClimateMarch network, where EAGLE achieves the lowest error distances and sizes of potential sources (\fref{fig:RealPath}(j) and (k)). While the OLS approach occasionally matches the size of EAGLE’s inferred source set, its F-score remains at 0, indicating inaccurate predictions, as illustrated in \fref{fig:RealPath}(l). 

Across all experimental scenarios, existing SOTA approaches achieve near-zero accuracy, as measured by the F-score, indicating their inability to infer true sources when confronted with real propagation paths. In contrast, the proposed EAGLE method achieves F-scores ranging from 0.1 to 0.4, a substantial improvement over competing models that fail completely (F-score equals 0). Remarkably, in networks with nearly 100,000 users, EAGLE is the only method capable of accurately identifying the true source with only 100 exposed observers, whereas others generate inferred sets exceeding 1,000 potential sources without containing the actual one.

These results demonstrate EAGLE's robustness and scalability in large-scale and uncertain environments. The consistent performance across real propagation data validates its ability to capture the intrinsic dynamics of social diffusion. Beyond technical accuracy, this capacity has critical implications for mitigating the spread of misinformation and managing public health risks, as it enables early detection and targeted intervention in vast and complex information ecosystems.

\section{Conclusion and Discussion} \label{sec.discussion}

This study introduces EAGLE, a comprehensive and generalizable framework for inferring information propagation sources in large-scale networks. This framework strategically integrates evolutionary optimization with generalized Bayesian inference, enabling accurate source identification with minimal observer input and no prior assumptions about propagation models or user attributes. Specifically, the discrete moth-flame optimization algorithm is utilized to optimally deploy observers across the network, maximizing the observation capability and enhancing situational awareness of limited observers. Unlike traditional probability functions, EAGLE employs belief functions to represent the likelihood of candidates being the source and incorporate uncertainty among indistinguishable candidates, allowing the model to reason effectively under incomplete or noisy observations. Using only the timing and direction of propagation from a small number of exposed observers, EAGLE performs early-stage source inference with minimal input, which is both practical and computationally efficient. Observational data are iteratively integrated through the generalized Bayesian inference model, where each observation is weighted by its credibility level. This approach mitigates the effects of noise and uncertainty, yielding accurate localization of potential sources.

Comprehensive experiments across synthetic and real-world networks under diverse propagation scenarios demonstrate that EAGLE consistently outperforms existing SOTA methods, that is, identifying the smallest candidate set that reliably contains the true source, and often pinpointing it exactly. Notably, when applied to Twitter datasets with real propagation paths, EAGLE achieves accurate source inference even under diverse, unmodeled, and stochastic user interaction patterns -- a feat not achieved by any comparative method.

By modeling the diffusion of misinformation and other contagion-like processes, EAGLE enables early, reliable, and interpretable source identification, thereby enhancing both theoretical understanding and practical management of propagation dynamics in complex systems. This capability provides decision-makers, organizations, and researchers with actionable insights to design timely countermeasures, thereby protecting public health, economic stability, and digital trust. Moreover, the generalized Bayesian framework underlying EAGLE opens new directions for research on data-driven inference under uncertainty in large, interconnected systems, suggesting broad applicability to domains such as cybersecurity, epidemiology, and power-grid resilience.

In the future, multi-source localization represents an important next step, as many misinformation campaigns originate simultaneously from multiple, loosely coordinated accounts. Developing techniques to identify these dispersed and dynamically evolving sources will be a key research direction.  
Moreover, adapting the framework to other domains, such as localizing faults in national power grids or identifying patient zero in spatially and temporally evolving epidemic networks, will require methodological advances in observer deployment, data integration, and spatiotemporal inference. These challenges offer clear directions for extending EAGLE's theoretical foundation and practical impact across diverse networked systems.

\section*{Declaration of competing interest}

The authors declare that they have no conflict of interest.

\section*{Acknowledgment}

This project was partially supported by the Economic and Social Research Council (ESRC), UK as part of the Centre for Doctoral Training in Data Analytics and Society, and The Alan Turing Institute’s Enrichment Scheme.

\section*{Declaration of generative AI and AI-assisted technologies in the manuscript preparation process}

During the preparation of this work, the authors used GPT-4 in order to improve language. After using this tool, the authors reviewed and edited the content as needed and take full responsibility for the content of the published article.

\bibliographystyle{unsrtnat}
\bibliography{MainBib}

\end{document}